\documentclass{agujournal}

\journalname{Space Weather}

\usepackage{graphicx}
\usepackage{amsmath}
\newcommand{\gt}{\textgreater}

\usepackage{rotating}

\usepackage{graphics}
\usepackage{epsfig}
\usepackage{url}
\usepackage{multirow}
\usepackage{times}
\usepackage{amsmath}
\usepackage{longtable}
\usepackage{url}
\usepackage{graphicx}
\usepackage[ansinew]{inputenc}
\usepackage{array}
\usepackage{parskip}
\usepackage{afterpage}
\usepackage{xr}

\begin{document}

%% ------------------------------------------------------------------------ %%
%  Title
%
% (A title should be specific, informative, and brief. Use
% abbreviations only if they are defined in the abstract. Titles that
% start with general keywords then specific terms are optimized in
% searches)
%
%% ------------------------------------------------------------------------ %%

% Example: \title{This is a test title}

\title{Electron intensity measurements by the Cluster/RAPID/IES instrument in Earth's radiation belts and ring current}

%% ------------------------------------------------------------------------ %%
%
%  AUTHORS AND AFFILIATIONS
%
%% ------------------------------------------------------------------------ %%

% Authors are individuals who have significantly contributed to the
% research and preparation of the article. Group authors are allowed, if
% each author in the group is separately identified in an appendix.)

% List authors by first name or initial followed by last name and
% separated by commas. Use \affil{} to number affiliations, and
% \thanks{} for author notes.
% Additional author notes should be indicated with \thanks{} (for
% example, for current addresses).

% Example: \authors{A. B. Author\affil{1}\thanks{Current address, Antartica}, B. C. Author\affil{2,3}, and D. E.
% Author\affil{3,4}\thanks{Also funded by Monsanto.}}

\authors{A. Smirnov\affil{1,2}, E. Kronberg\affil{2,1}, F. Latallerie\affil{1,2}, P. W. Daly\affil{2}, N. Aseev\affil{3}, Y. Shprits\affil{3,4}, A. Kellerman\affil{4}, S. Kasahara\affil{5}, D. Turner\affil{6}, M.G.G.T. Taylor\affil{7}, S.Yokota\affil{8}, K. Keika\affil{5}, T. Hori\affil{9}}

\affiliation{1}{Department of Earth and Environmental Sciences, Ludwig Maximilians University, Munich, Germany.}
\affiliation{2}{Max Planck Institute for Solar System Research, G\"{o}ttingen, Germany.}
\affiliation{3}{Helmholtz Centre Potsdam GFZ German Research Centre for Geosciences and Institute of Physics and Astronomy, University of Potsdam, Potsdam, Germany.}
\affiliation{4}{Department of Earth Planetary and Space Sciences, University of California, Los Angeles, California, USA.}
\affiliation{5}{The Aerospace Corporation, El Segundo, California, USA.}
\affiliation{6}{Department of Earth and Planetary Science, Graduate School of Science, The University of Tokyo, Tokyo, Japan.}
\affiliation{7}{ESA/ESTEC, Noordwijk, Netherlands}
\affiliation{8}{Department of Earth and Space Science, Graduate School of Science, Osaka University, Osaka, Japan.}
\affiliation{9}{Institute for Space-Earth Environmental Research, Nagoya University, Aichi, Japan.}

%(repeat as many times as is necessary)

%% Corresponding Author:
% Corresponding author mailing address and e-mail address:

% (include name and email addresses of the corresponding author.  More
% than one corresponding author is allowed in this LaTeX file and for
% publication; but only one corresponding author is allowed in our
% editorial system.)

% Example: \correspondingauthor{First and Last Name}{email@address.edu}

\correspondingauthor{Artem Smirnov}{arsmirnov95@gmail.com}

%% Keypoints, final entry on title page.

% Example:
% \begin{keypoints}
% \item	List up to three key points (at least one is required)
% \item	Key Points summarize the main points and conclusions of the article
% \item	Each must be 100 characters or less with no special characters or punctuation
% \end{keypoints}

%  List up to three key points (at least one is required)
%  Key Points summarize the main points and conclusions of the article
%  Each must be 100 characters or less with no special characters or punctuation

\begin{keypoints}
\item We present two algorithms for background correction of RAPID electron measurements at energies 40-400 keV
\item Corrected RAPID data coincide well with Van Allen Probes and Arase electron measurements
\item The generalized relationship between IES electron flux intensities and solar wind dynamic pressure was obtained
\end{keypoints}

%% ------------------------------------------------------------------------ %%
%
%  ABSTRACT
%
% A good abstract will begin with a short description of the problem
% being addressed, briefly describe the new data or analyses, then
% briefly states the main conclusion(s) and how they are supported and
% uncertainties.
%% ------------------------------------------------------------------------ %%

%% \begin{abstract} starts the second page

\begin{abstract}
The Cluster mission, launched in 2000, has produced a large database of electron flux intensity measurements in the Earth's magnetosphere by the Research with Adaptive Particle Imaging Detector (RAPID)/ Imaging Electron Spectrometer (IES) instrument. However, due to background contamination of the data with high-energy electrons (\gt400 keV) and inner-zone protons (230-630 keV) in the radiation belts and ring current, the data have been rarely used for inner-magnetospheric science. The current paper presents two algorithms for background correction. The first algorithm is based on the empirical contamination percentages by both protons and electrons. The second algorithm uses simultaneous proton observations. The efficiencies of these algorithms are demonstrated by comparison of the corrected Cluster/RAPID/IES data with Van Allen Probes/Magnetic Electron Ion Spectrometer (MagEIS) measurements for 2012-2015. Both techniques improved the IES electron data in the radiation belts and ring current.Yearly averaged flux intensities of the two missions show the ratio of measurements close to 1. IES corrected measurements were also compared with Arase Medium-Energy Particle Experiments-Electron Analyzer (MEP-e) electron data during two conjunction periods in 2017 and also exhibited ratio close to 1. We demonstrate a scientific application of the corrected IES electron data analyzing its evolution during solar cycle. Spin-averaged yearly mean IES electron intensities in the outer belt for energies 40-400 keV at L-shell between 4 and 6 showed high positive correlation with AE index and solar wind dynamic pressure during 2001- 2016. Relationship between solar wind dynamic pressure and IES electron measurements in the outer radiation belt was derived as a uniform linear-logarithmic equation.
\end{abstract}

%% ------------------------------------------------------------------------ %%
%
%  TEXT
%
%% ------------------------------------------------------------------------ %%

%%% Suggested section heads:
% \section{Introduction}
%
% The main text should start with an introduction. Except for short
% manuscripts (such as comments and replies), the text should be divided
% into sections, each with its own heading.

% Headings should be sentence fragments and do not begin with a
% lowercase letter or number. Examples of good headings are:

% \section{Materials and Methods}
% Here is text on Materials and Methods.
%
% \subsection{A descriptive heading about methods}
% More about Methods.
%
% \section{Data} (Or section title might be a descriptive heading about data)
%
% \section{Results} (Or section title might be a descriptive heading about the
% results)
%
% \section{Conclusions}

\section{Introduction}

The Van Allen radiation belts are zones of charged energetic particles, mainly electrons and protons, trapped within the magnetic field of the Earth. Protons form a single radiation belt with maximum flux between L values of 3 to 4 \citep{Ganushkina2011}. Electron radiation belts exhibit a two-zone structure: the inner radiation belt lies between L values from 1 to 3 and has a peak of electron flux around L values of 2 to 3 \citep{Lyons1972}. The outer radiation belt lies within L values of 3 to 7, exhibiting the maximum electron flux intensity at L=4-5 \citep{Kellerman2014}. We note that a third electron belt has also been recently reported \citep{Baker2013}.\\

The radiation belts were discovered by the array of Explorer satellite missions \citep[e.g.,][]{Williams1960}. The Cluster mission \citep[e.g.,][]{Escoubet1997, Wilken2001} has been investigating plasma phenomena in the magnetosphere and near Earth environment for more than 15 years. Its orbit has evolved from its initial near polar 4x19 $R_{E}$ but has systematically passed through the radiation belt regions. Energetic electron flux measurements have been made by RAPID/IES detector. However, electron flux measurements of the IES detector in range 30-400 keV are susceptible to background contamination effects in the radiation belts region, which can cause significant errors in measurements and thus have made it challenging to use RAPID/IES data around the pericentre of its orbit. Background contamination can be a result of penetrating radiation and energetic particles, amongst which perhaps the most harmful are the  contamination effects caused by protons with energies \gt 100 keV on the corresponding L values of 3 to 4, and by high-energy electrons ( \gt 400 keV) on L values from 4 to 6. \\

In the current paper we describe the IES electron flux measurement corrections based on two methods. Firstly, we will correct for the background contamination using the techniques described in \citep{Kronberg2016}. Using the radiation transport code Geant4 \citep{Agostinelli2003, Allison2006}, \citet{Kronberg2016} analyzed the efficiency of IES instrument shielding against the penetrating particles by modeling the geometric configuration of IES detector and bombarding it with proton and electron fluxes on the angles from 0$^{\circ}$ to 90$^{\circ}$ with a step of 1$^{\circ}$. In the course of the experiment, the authors derived the percentages of contamination of the IES detector for the energy channels 1-6 for L values from 1 to 9. In order to verify the results of the data filtering, we compare the corrected data to the Van Allen Probes/MagEIS \citep{Blake2013} electron flux measurements. Secondly, we use proton flux observations from Van Allen Probes/ MagEIS for background correction. As demonstrated by \citet{Kronberg2016}, protons of energies $>$230 keV are considered to contaminate the IES measurements. Using Van Allen Probes proton flux measurements with energies in range 200-600 keV, we substract their flux intensities from RAPID/IES electron flux observations and minimize the difference between Van Allen Probes and Cluster observations. \citet{Kronberg2016} compared the corrected IES data with MagEIS for a period of 3 months and used only energies from 240 to 400 keV, while in this study we use 5 years of MagEIS data and energies from 40 to 400 keV. \\

The paper has 6 parts, including this introductory section. In section 2, we describe the data used in this study. Section 3 presents the background correction algorithm based on empirical contamination coefficients. Section 4 describes the correction technique using the RBSP-B/MagEIS proton measurements. Section 5 shows the comparison with Arase MEP-e electron measurements during two conjunction periods. Section 6 is concerned with scientific applications of the corrected RAPID/IES electron data, particularly for analysis of the solar cycle dependence of the electron intensity variation. Section 7 draws upon the entire paper and gives the conclusions.

\section{Data}

Launched in 2000, the Cluster constellation consists of four identical spacecraft following the elliptical polar orbits with an initial perigee at about 4 $R_E$ and apogee at 19.6 $R_E$ \citep{Escoubet1997}. Electrons with energies 30-400 keV are measured with the RAPID/IES detector on board the Cluster mission. IES is a solid-state silicon sensor, which consists of three acceptance "pin-hole" systems, each dividing a 60$^{\circ}$ segment into 3 angular intervals. Three of such detectors are put in configuration which provides electron flux measurements over a 180$^{\circ}$ fan \citep{Wilken1997}. In our study we use spacecraft 4 (Tango).\\

The Van Allen Probes (RBSP) constellation, launched in 2012, consists of two  spacecraft orbiting the Earth on near-equatorial orbit with apogees at L=6 and perigees at altitude of 700 km. Magnetic Electron Ion Spectrometer (MagEIS) instruments aboard each of the two RBSP spacecraft (RBSP-a and RBSP-b) measure electron fluxes on one low-energy unit (20-240 keV), two medium-energy units (80-1200 keV), and a high-energy unit (800-4800 keV). The instruments also contain a proton telescope, measuring protons with energies 55 keV- 20 MeV \citep{Blake2013}. The corresponding energy channels for RAPID/IES and MagEIS detectors are shown in Tables 1 and 2, respectively. \\

The Arase (ERG) mission \citep{Miyoshi2018} consists of a spacecraft launched on the 20th of December 2016 to explore the Earth radiation belt region. The spacecraft altitude is 440 km in perigee and 32,000 km in apogee with an inclination of approximately 31$^{\circ}$. The energy and direction of incoming electrons are measured by the Medium-Energy Particle Experiments-Electron Analyzer (MEP-e) \citep{Kasahara2018}.\\	

From Tables 1 and 2 it is evident that IES channels 1-4 coincide fairly well with channels 0-3 of MagEIS, as the difference between the lower thresholds for the two missions is within 10\% error. Hence, we can compare their data without adjusting these channels to each other. On the other hand, MagEIS channels 4, 5 and 6 coincide with IES channel 5, and MagEIS channels 7 and 8 - with RAPID channel 6. In order to compare the data of MagEIS and IES, we need to recalculate the electron flux intensities on the specified MagEIS channels. For that we apply the formula following \citep{Kronberg2012}: \\
\begin{center}
	$I'_{j} = \dfrac{\sum_{i=1}^{n} I_{i}*(E_{i+1}-E_{i})}{E_{max}-E_{min}}$
	
\end{center}

where $ I'_{j}$ is the equivalent MagEIS electron intensity for the RAPID/IES energy channel $j$, $ I_{i}$ is the MagEIS intensity for channel $ i $, and $E_i$ are the corresponding lower thresholds of energy channels, $E_{max}$ and $E_{min}$ are the maximal and minimal energies for the energy interval $ I_{j}$. \\ 

We apply two methods of the background correction. First, we correct for electron and proton contamination using coefficients, obtained by the empirical experiment. Second, we subtract the MagEIS proton measurements from the IES electron observations. After that, we compare the derived data with the corrected MagEIS electron measurements \citep{Claudepierre2015} and minimize their difference in order to assess the percentage of protons that reach the IES detector. The subsequent analysis will be based on RBSP-B data.\\

%------------------------------------------------

\section{Background correction using empirical contamination coefficients}

\citet{Kronberg2016} showed that one of the main sources of errors in the electron flux measurements are the electrons with energies higher than 400 keV, as they can deposit energies less than 400 keV in the detector. Their influence becomes quite substantial on the highest energy channels. It was also shown that protons must possess the energy of at least 230 keV in order to influence the measurements. The foil, the dead layer in front of the detector and the electronic noise define the lowest energy of incoming protons being 200 keV, and together with the lower limit of electronics aboard (32 keV), they must have at least 230 keV to be detected by the IES sensor. Protons with energies $>$630 keV are successfully cut off by electrons, so we only consider protons of energies from 230 to 630 keV as contaminants. Contamination by protons is especially strong at L values of 3-4, as proton flux peaks at these L-shells. Combining the percentages of contamination by protons and electrons at different L values for energy channels from 1 to 6, \citet{Kronberg2016} derived the percentages of total contamination, shown in Table 3. \\

In order to filter the IES electron flux measurements based on the contamination coefficients, we subtract the part of the flux intensity attributed to the influence of background contamination: \\
\begin{center}
	$I_{electrons} = I_{measured}*(1-\frac{p}{100})$, \\ 
\end{center}

where $ I_{measured} $ is the total measured electron flux intensity, p - percent of contamination from the Table 3, $ I_{electrons} $ - the filtered electron intensity. \\

Figure 1 presents the comparison between the spin-averaged yearly mean RAPID/IES electron measurements, corrected by the above-described method, and equivalent RBSP-B/MagEIS electron measurements for years 2012-2015. The choice of years for comparison was due to the fact that Van Allen Probes mission was launched in 2012, and already in 2016 the Cluster and Van Allen Probes mission had no intercept in altitudes on which they were flying. In 2016 the Cluster mission did not fly on L-shells lower than 5.8, which was the apogee for RBSP mission. The ratio between corrected RAPID and MagEIS electron intensities is close to 1, as the median values of the ratio for years 2012-2015 equals to 1.0 in all cases, while the less stable characteristic, the mean value, stays very close to 1 (Table 4 presents mean and median ratio values over the whole Energy channel $*$ L matrix). It can be concluded that RAPID/IES electron intensities observations are in very good agreement with those of MagEIS, considering that the two instruments have different measuring techniques, as well as different calibration algorithms used \citep{Claudepierre2015, Daly2010}. We note that although there is significant L shell overlap, the two missions have very different orbits, where Van Allen has an equatorial orbit, and Cluster is more polar, spending a majority of the time at higher latitudes. \\

The overall effect of the correction algorithm is evident in Figure 2, which shows the ratio between the spin-averaged yearly mean IES electron intensities and the uncorrected measurements for years 2001-2016. The ratio is especially high on higher channels at L-shells around 3-4, where there is a peak in proton intensities due to the proton radiation belt.\\

\section{Background correction using MagEIS proton measurements}

In order to correct IES data for proton contamination, we obtain the mean values of spin-averaged fluxes for 6 channels, which are corrected for electron contamination using the coefficients obtained by \citep{Kronberg2016}, and collect them in single energy channel bins in 0.2 L-shell increments. After that, the yearly averaged proton intensities values for energies 230-630 keV are calculated for every 0.2$L$ bin and subtracted from the IES data. We note again that the two instruments are not identical and therefore have different calibration and measurement responses. RAPID/IES electron measurements can also be susceptible to cosmic ray background noise and also electron sensor degradation since 2007 \citep{Kronberg2016b}. Hence, we have to use a functional combining the above-mentioned effects. There are several ways of constructing such functional (1), for instance, we can use the one derived by \citet{Kronberg2016}:\\
\begin{equation}
median \{{\log_{10}Y - [b\log_{10}(X\circ M) - \log_{10}c \cdot \log_{10}P - \log_{10}a]\}}^2 \rightarrow min,
\label{eq1}
\end{equation}
where $Y$ is the MagEIS electron data, $X$ is IES data, $M$ is the correction matrix for electron contamination derived by Geant4 observations, $X\circ M$ is the Hadamard (element-wise) product of two matrices X and M, $P$ is the matrix of proton intensities observed by RBSP-B/MagEIS instrument at energies 230-630 keV. Parameters $a$, $b$ and $c$ are used for optimization, and they represent background noise, sensor degradation, and the optimal percentage of protons that reach the detector, respectively. The authors used the logarithmic values of all parameters, due to the fact that intensity values can vary drastically, up to several orders of magnitude, and because the energy spectra follow exponential power laws in this range \citep{Cayton1989}.\\

Such a function was applied by \citep{Kronberg2016} to the IES electron measurements at channel 6 during 245-366 DOY 2012 and obtained the best fit for $a = 0$, $b = 1$, and $0.55<c<1$. In the current study we apply a similar approach to all 6 energy channels, but over a much longer time period to provide better statistics (2012-2015). We minimize the functional using the differential evolution method, implemented in SciPy optimization package. In the first optimization approach, we try to use fixed values of $\log_{10}a=0$ and $\log_{10}b=1$, setting $c$ to be the optimization parameter after \citet{Kronberg2016}. In the second approach, we set $a$, $b$ and $c$ with the corresponding ranges of $[0;0.5]$, $[0;1]$, and $[0;1]$. The results are shown in Figure 3. It is worth mentioning that the first optimization approach with fixed $a$ and $b$ leads to the value of $c=0.99$. However, such an approach has several drawbacks. Electron intensities on higher channels are substantially lower than on lower energy channels. Consequently, when subtracting proton intensities, we get negative values of intensities for the large parts of higher energy channels data, which is physically incorrect. \\

In the following we describe our third approach. Here the IES electron observations must be close to those of RBSP-B/MagEIS instrument. Therefore, their ratio must be close to 1. In order to have such effect, we construct the following functional (2):
\begin{equation}
\sum\limits_{i,j} {\left( \dfrac{\log_{10}b \cdot \log_{10}(X_{ij}\cdot M_{ij}) - \log_{10}c \cdot \log_{10}P_{ij} - \log_{10}a}{\log_{10}Y_{ij}} - 1  \right)}^2 \rightarrow min,
\label{eq2}
\end{equation}
where the notations are as in functional (1).
The results of application of both functionals (1) and (2) are shown in Figure 3.

Table 5 presents the numerical results of optimization procedure for Functionals (1) and (2) in 2013, with $f$ being the value of the corresponding functional, $p$ - the mean value of the IES/MagEIS intensities ratio, and $m$ - the median value of the ratio over all L and E. Mean and median values of the electron intensities ratio are higher for Functional (2). However, this effect is attributed to negative intensity values in higher energy channels (see Figure 3) after subtraction of high proton intensities. On the other hand, we observe almost identical ratio values for two functionals on L-shells 2-3. Taking into account the fact that we compare measurements of two different instruments, the observed intensity ratio of 2.4 is satisfying for our purposes. The mean and median values for functional (2) are shown in Table 6. For years 2013-2015, the ratio values were close to 1. From Table 6, one can conclude that the optimal values for background noise, sensor degradation and percent of protons reaching the detector (a, b and c values) will be 10, 100 and 6, respectively.\\

The subtraction of proton intensities from the IES electron measurements proved to be a reliable background correction mechanism, as it brings the electron intensities ratio between the corrected RBSP-B/MagEIS and IES measurements close to 1, and the values of optimization parameters $a$ and $b$ coincide fairly well with those derived by \citet{Kronberg2016}. However, the value of $c$ is different: according to \citet{Kronberg2016}, the optimal value of $c$ is between 0.55 and 1, while in the current study we derived values of $c$ smaller than 0.1. This might be due to different datasets used in two studies. \citet{Kronberg2016} analyzed one energetic channel and $\approx 3$ months of observations. Here we subtracted proton fluxes at 230 to 630 keV from all six channels for much longer time period (2012-2015).\\

Both background correction mechanisms, based on empirical contamination coefficients and on MagEIS proton measurements, are suitable for electron intensities data correction. However, in spite of the fact that the first method is based on a statistical model, it shows better results and is more convenient to use due to the fact that it uses only contamination coefficients obtained by bombarding the same instrument configuration with energetic particles. The second method uses the combination of the electron contamination coefficients and MagEIS proton measurements (or other available proton measurements), which can lead to further uncertainties caused by differences between the two instruments. 

\section{Comparison of RAPID/IES and Arase MEP-e electron intensities}

We compare the electron intensities measured by the Cluster 4 RAPID/IES and the Arase MEP-e instruments within the energy range from 68 keV to 96 keV in radiation belts in 2017. We searched for time periods when the two missions were on the same drift path. In this case it is not necessary for the spacecrafts to be physically close to each other. To do this, we define a conjunction period by the following criteria: (1) The spacecrafts separation must be less than 0.2 Earth radii. (2) The satellites must also be close to the magnetic equator. For this condition to be satisfied, we calculate the ratio between the local and the equatorial magnetic field magnitudes. This ratio must be more than 0.8 and less than 1.2. The absolute value of the ratios difference for the Cluster 4 and the Arase spacecrafts must also be less than 0.2. The values for the magnetic equator are based on the Olson-Pfitzer Quiet model \citep{Olson1974}. (3) Therefore, the magnetosphere must be quiet during the conjunction periods. Namely, a conjunction period is always when the Kp index is less than 3 for at least two days before the conjunction. We get the Kp indexes from the International Association of Geomagnetism and Aeronomy [http://isgi.unistra.fr/indices\_kp.php]. (4) The spacecraft measurements are allowed to be correlated within a time interval of 3 hours.
To match the MEP-e and RAPID/IES energy ranges, the intensities are recalculated using the method presented in the Section 2 (Equation 1). The intensities measured by the Cluster 4 are also corrected with the background correction using empirical contamination coefficients (from \citet{Kronberg2016}, see Table 3).\\

We found two time periods satisfying the conjunction criteria. They are represented by the two clusters of points in Figure 5. The one corresponding to the lowest intensities happened on April 18, 2017. During this conjunction, the mean position vectors in the GSE system were [ -4.31 , -3.39 , 2.58 ] for the Arase satellite (L-shell 6.8) and [ -4.27 , 4.22 , 1.73 ] for the Cluster 4 (L-shell 6.5). The upper cluster was measured on April 27, 2017 when the Arase satellite and Cluster 4 where orbiting at the averaged positions [-4.63, -2.68, 2.86] (on L-shell 6.6) and [-3.57, 4.89, 1.61] on L-shell 6.6, respectively.
The slope of the linear regression close to one shows that the electron intensities are similar.

\section{Application to solar cycle analysis}
The Cluster mission has been operating since 2001, therefore it can provide a rich long term database. The mission's lifespan covers almost two solar cycles: solar cycle 23 (1996-2008) and solar cycle 24 (2008-2019). In this section we demonstrate the results of solar cycle monitoring for years 2001- 2016. IES electron flux intensities, corrected using the contamination coefficients from Table 3, are presented in Figure 6.\\

The 10.7 cm solar radio flux ($F_{10.7}$) is used as a simple activity level indicator or as a proxy for the solar emissions \citep{Tapping2013}. The index takes into account three different mechanisms of emission: thermal-free emission from the chromosphere and corona, solar activity associated with Coronal Mass Ejections (CMEs) which trigger magnetic storms and may lead to significant enhancement in radiation belt fluxes, and also nonthermal emissions \citep{Tapping2013}. In Figure 7 (A), we demonstrate 27-day averaged and yearly averaged $F_{10.7}$ index for 2001-2016. It exhibits a pick in 2001 with $F_{10.7}=128.45$, the year of solar maximum in cycle 23. After that, the index values started to decrease towards a solar cycle 23-24 minimum in 2008, which was the historical minimum of solar activity with yearly averaged number of $F_{10.7}=69$ and 265 spotless days [data source: NOAA Space Weather Prediction Center, 

https://www.ngdc.noaa.gov/stp/solar/solar-indices.html].\\ 

Auroral electrojet (AE) index was introduced by \citet{Davis1966} as a difference between the upper and lower envelopes of the superposed H component magnetic variations from 12 magnetic observatories situated at 61$^{\circ}$ - 70$^{\circ}$ latitude in the Northern Hemisphere \citep{Kamide1983}. The greater values of AE index stand for the substorm activity\citep{Kamide1983}. In Figure 7 (B), we show the 27-day averaged and yearly averaged values of AE index in 2001- 2016. It has a clear pick in 2003, during the declining phase of solar cycle 23, and a global minimum in 2008-2009, during the solar cycle 23-24 minimum.\\

In Figure 7 (C) we demonstrate 27-day and yearly averaged solar wind dynamic pressure, which is one of the most important solar wind parameters for determining effects on the magnetosphere. Increase (or decrease) in the solar wind dynamic pressure will compress (decompress) the magnetosphere and cause variation in the strength of magnetic field. From Figure 7 (B and C) one can see that the dynamic pressure exhibits the pattern very similar to that of AE index: it has a strong maximum in 2003, a global minimum in 2008 with the gradual following increase up to 2015.\\

In order to compare these parameters with RAPID/IES measurements, in Figure 7 (D) we also demonstrate spin-averaged yearly mean electron flux intensities [in keV$^{-1}$sr$^{-1}$cm$^{-2}$s$^{-1}$] for channels 1-6, averaged over $L=4-6$. The curves for all six channels exhibit roughly the same behavior with the indices described above: one can observe a pick in 2003 (during the declining phase of solar cycle 23) with gradual decrease towards 2008-2009 during $Smin$ of cycle 23-24, followed by an increase in intensities up to 2015. \\ 		  

In order to obtain the statistical relation between electron flux intensities measured by RAPID/IES detector and yearly averaged solar wind dynamic pressure shown in Figure 7, we perform linear regression analysis for energy channels 1-6, shown in Figure 8. We normalize our data using the following formula:
\begin{center}
	$X'=\dfrac{X-X_{min}}{X_{max}-X_{min}}$,
\end{center}
where $X'$ stands for the normalized data, $X$- not normalized data, $X_{min}$ and $X_{max}$ represent the maximum and minimum value of the averaged data.\\		

In Figure 8, one can observe the moderate uphill (positive) linear relationship. All four plots exhibit an intercept around 0.33, and the slope of $\approx 0.5$. The more precise results are presented in Table 8. In the linear regression analysis we use solar wind dynamic pressure as a predictor variable, and the electron flux intensities for channels 1-6 as response variables. The Pearson linear correlation coefficient R was $\approx 0.5$ for channels 2, 4 and 5 and lower (0.4) for channel 3, due to the greater variance in the data.\\

Linear trend equations for IES electron flux intensities are presented in Table 9. We can derive a generalized relationship between solar wind pressure and IES electron flux intensity measurements at energies from 40 to 400 keV in the form:
\begin{center}
	$\log_{10}\textbf{y}=\textbf{0.35}(\pm0.1) + \textbf{0.5}(\pm0.18)\textbf{x},$
\end{center}
where $x$ stands for normalized flow pressure, and $y$ represents normalized electron fluxes measured by the IES detector on board Cluster mission at L-shells 4 to 6. \\

These findings coincide well with previous studies \citep[e.g.,][]{Baker2001,Li2001,Li2006}. \citet{Baker2001} concluded that during the declining phase of solar cycle approaching sunspot minimum, the recurrent high speed solar wind streams are driving recurrent magnetic storms and enhancing radiation belts electrons. \citet{Li2001} supported that conclusion and stated that during the ascending phase of solar cycle, the enhancement in radiation belts' electrons is mainly caused by coronal mass ejections (CMEs). Geoeffective CMEs do not occur as often as the high speed solar wind streams. Therefore, electron flux intensities in the outer belt are higher during the descending phase of solar cycle. \citet{Li2006} analyzed monthly window-averaged fluxes of 2-6 MeV electrons from SAMPEX data and demonstrated that the outer radiation belt was most intense during the descending phase of the sunspot cycle, and weakest during the sunspot minimum, but also that the outer belt was not in its most active state when approaching the solar maximum. In the present study we observed the maximum of electron fluxes in year 2003, i.e. during the descending phase of solar cycle 23 followed by the decrease in intensities up to the minimum of solar cycle 23-24 in 2008-2009. After that the intensities started growing up to 2012, having a slight decrease afterwards with the local minimum in 2014 (a maximum of solar cycle 24). It seems that the IES electron flux measurements correlate very well in the time domain with AE index and dynamic flow pressure (Figure 7 B, C), while for years 2012-2016 all three parameters (electron fluxes, AE index, and dynamic pressure) seem to be anti-correlated with $F_{10.7}$. \\

\citet{Miyoshi2004} analyzed the variation in electron flux intensities for energies $>$30 keV using long-term SAMPEX observations. The study revealed two main results: (1) significant electron flux variations with the solar cycle, and (2) location of the electron flux maximum in the outer belt shifted to lower L-shells during periods of higher magnetic activity, and vice versa, during periods of low magnetic activity the electron flux maximum was shifted to higher L-values, which coincides well with the present study. In Figure 6, one can observe that the electron flux maximum in the outer belt for 2009 (which was a year of low solar activity) was situated on $L\approx 6-7$, while in 2001 (maximum of solar cycle 23), it was situated on $L\approx 4-6$. Our study agrees with the work by \citet{Miyoshi2004}, which shows that the electron fluxes are increased during the active solar periods (especially at the declining phase) and decreased during the quiet periods. However, conclusions by \citet{Miyoshi2004} about anti-correlation for electrons at energies $>$300 keV with solar activity index at L-shells from 5 to 6 during solar cycles 21-23 do not coincide with the current findings. In our case the fluxes do not show any correlation with the solar activity index. \\

\section{Conclusions}
The electron data measured by the IES detector on board Cluster II mission is subject to contamination at energy interval 40-450 keV (energy channels 1-6). The main sources of contamination are inner zone protons (L-shells from 3 to 4) with energies 230-630 keV, and the high-energy electrons ($>$400 keV) at L-shells 4-6.\\
The current paper suggests two algorithms for background correction of RAPID/IES electron measurements. The first correction technique is based on application of experimentally obtained contamination coefficients after \citep{Kronberg2016}. The second method uses RBSP-B/MagEIS proton measurements with energies 230-630 keV. Both methods proved to be reliable in removing the contamination by high-energy electrons ($>$100 keV) and inner-belts protons (230-630 keV). However, it is more convenient to use the method I, as it does not require other observations.\\
The corrected data can be useful in many scientific applications. In the present study we analyzed spin-averaged yearly mean electron flux intensity variation along solar cycle 23-24. The study revealed that outer belt electrons exhibit the pattern very close to that of AE index and solar dynamic wind pressure. The linear regression analysis was applied in order to obtain regression models for logarithms of electron intensities on channels 1-6 versus flow dynamic pressure. The model was approximated by the linear fit, which leads to a generalized equation of relationship between solar wind dynamic pressure and IES electron flux intensities at energies from 40 keV to 400 keV at L-shells 4 to 6. These results can have wide space weather applications.

\section{= enter section title =}

\acknowledgments
 We acknowledge the
 Deutsches Zentrum f{\"u}r Luft und Raumfahrt (DLR) for supporting the RAPID
 instrument at MPS under grant 50 OC 1602. The Cluster data can be found at
 CSA Archive: http://www.cosmos.esa.int/web/csa/. We acknowledge RBSP ECT team
 for the MagEIS data. They can be found at http://www.rbsp-ect.lanl.gov/.  \\
 Science data of the ERG (Arase) satellite were obtained from the ERG
 Science Center operated by ISAS/JAXA and ISEE/Nagoya
 University (https://ergsc.isee.nagoya-u.ac.jp/index.shtml.en). The
 Arase satellite data will be publicly available via ERG Science Center
 on a project-agreed schedule. The present study analyzed the MEP-e v01\_00 data. Part of the work of TH was done at the ERG-Science Center operated by
 ISAS/JAXA and ISEE/Nagoya University.

%% ------------------------------------------------------------------------ %%
%% Citations

% Please use ONLY \citet and \citep for reference citations.
% DO NOT use other cite commands (e.g., \cite, \citeyear, \nocite, \citealp, etc.).

%% Example \citet and \citep:
%  ...as shown by \citet{Boug10}, \citet{Buiz07}, \citet{Fra10},
%  \citet{Ghel00}, and \citet{Leit74}.

%  ...as shown by \citep{Boug10}, \citep{Buiz07}, \citep{Fra10},
%  \citep{Ghel00, Leit74}.

%  ...has been shown \citep [e.g.,][]{Boug10,Buiz07,Fra10}.

%%  REFERENCE LIST AND TEXT CITATIONS
%
% Either type in your references using
%
\bibliographystyle{agu04}
 
%
%%%%%%%%%%%%%%%%%%%%%%%%%%%%%%%%%%%%%%%%%%%%%%%
% Or, to use BibTeX:
%
% Follow these steps
%
% 1. Type in \bibliography{<name of your .bib file>}
%    Run LaTeX on your LaTeX file.
%
% 2. Run BiBTeX on your LaTeX file.
%
% 3. Open the new .bbl file containing the reference list and
%   copy all the contents into your LaTeX file here.
%
% 4. Run LaTeX on your new file which will produce the citations.
%
% AGU does not want a .bib or a .bbl file. Please copy in the contents of your .bbl file here.

%% After you run BibTeX, Copy in the contents of the .bbl file here:

%%%%%%%%%%%%%%%%%%%%%%%%%%%%%%%%%%%%%%%%%%%%%%%%%%%%%%%%%%%%%%%%%%%%%
% Track Changes:
% To add words, \added{<word added>}
% To delete words, \deleted{<word deleted>}
% To replace words, \replaced{<word to be replaced>}{<replacement word>}
% To explain why change was made: \explain{<explanation>} This will put
% a comment into the right margin.

%%%%%%%%%%%%%%%%%%%%%%%%%%%%%%%%%%%%%%%%%%%%%%%%%%%%%%%%%%%%%%%%%%%%%
% At the end of the document, use \listofchanges, which will list the
% changes and the page and line number where the change was made.

% When final version, \listofchanges will not produce anything,
% \added{<word or words>} word will be printed, \deleted{<word or words} will take away the word,
% \replaced{<delete this word>}{<replace with this word>} will print only the replacement word.
%  In the final version, \explain will not print anything.
%%%%%%%%%%%%%%%%%%%%%%%%%%%%%%%%%%%%%%%%%%%%%%%%%%%%%%%%%%%%%%%%%%%%%
\begin{sidewaysfigure}
	\centering
	\includegraphics[width=\textwidth]{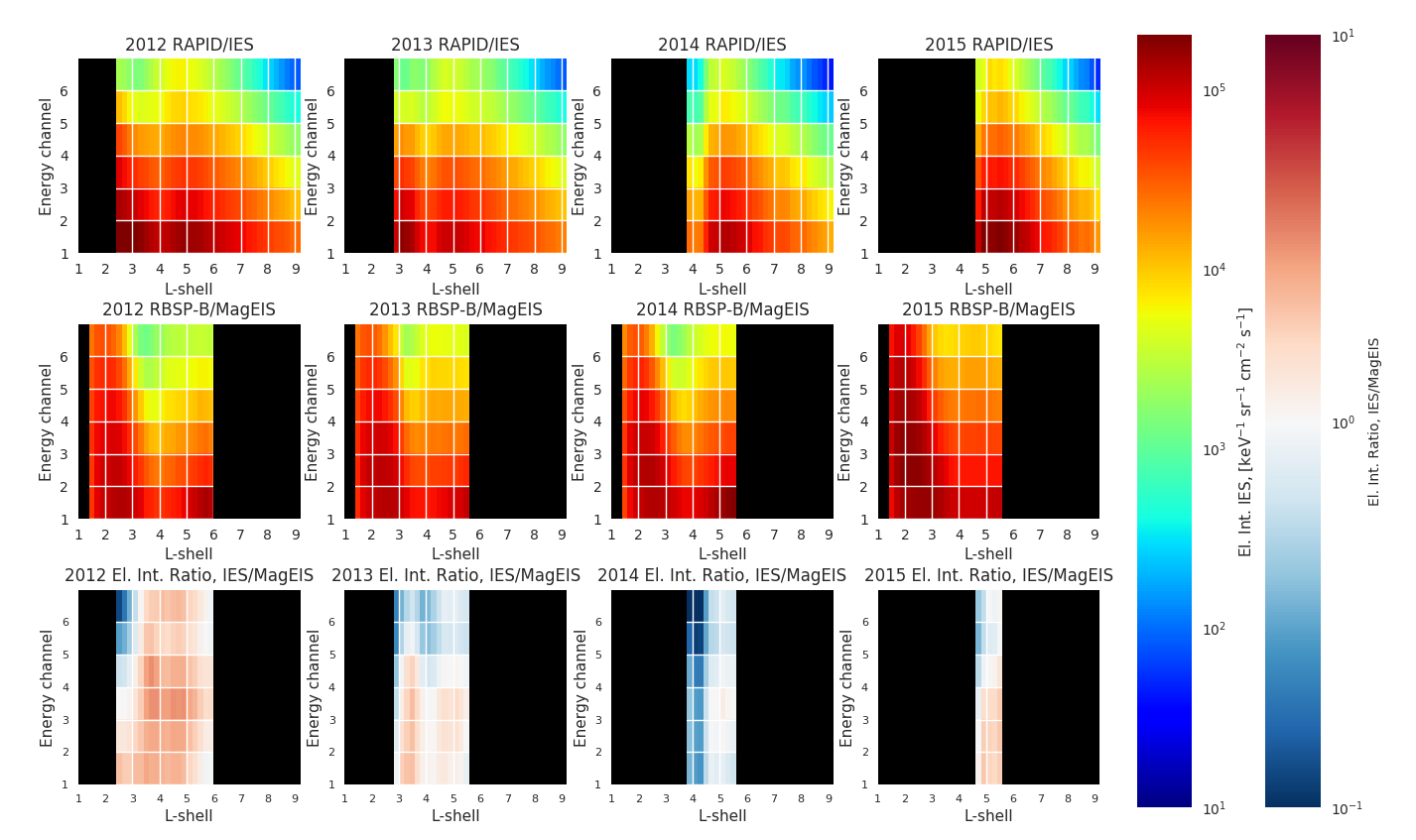}
	\caption{Corrected Cluster/RAPID/IES electron data, RBSP-B/ECT/MagEIS electron data, and their ratio for 2012-2015. Areas with no data are plotted in black}
	\label{2012_2015}
\end{sidewaysfigure}

%\newpage
\begin{sidewaysfigure}
	\centering
	\includegraphics[width=\textwidth]{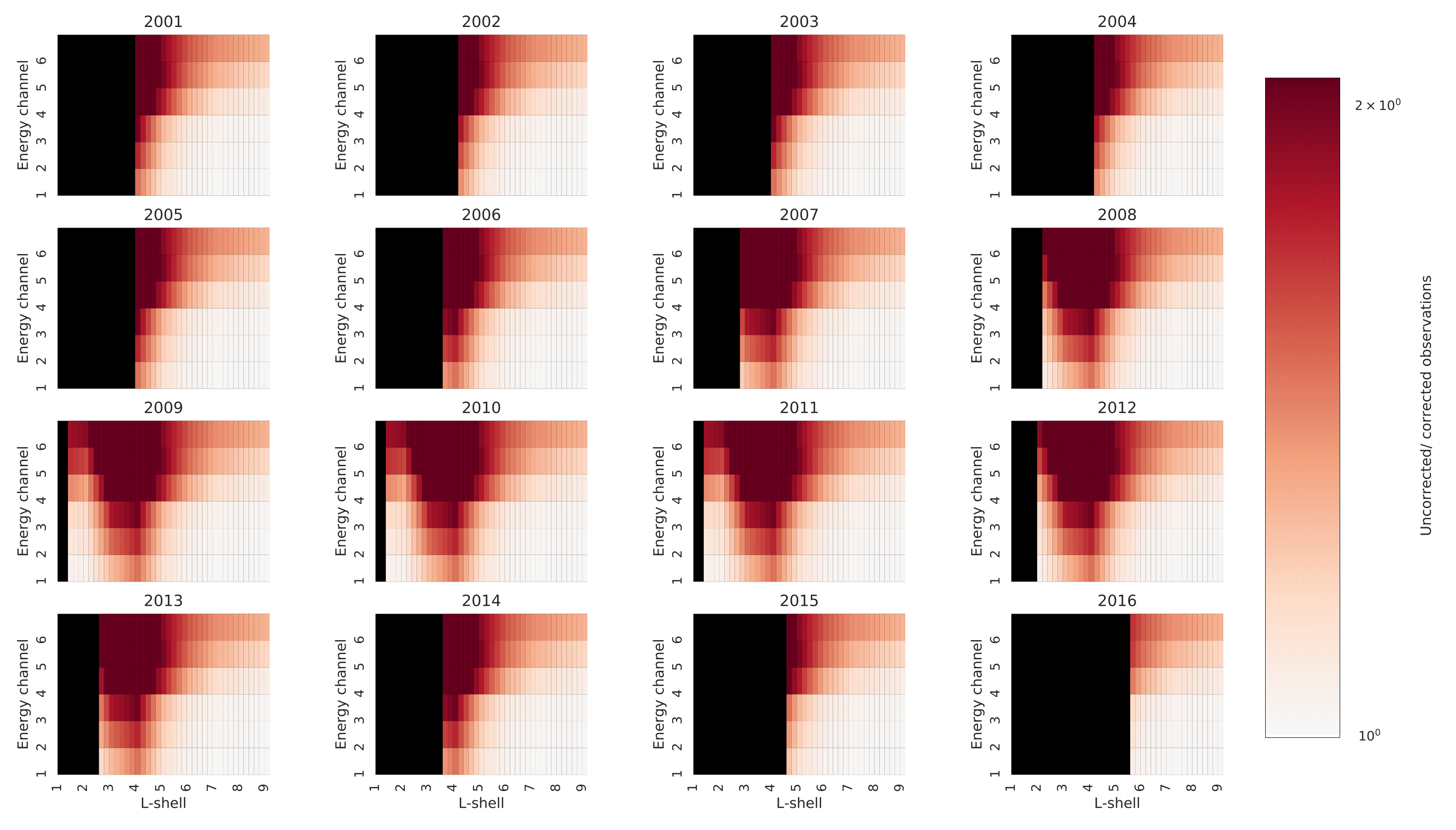}
	\caption{Ratio of uncorrected and corrected (using contamination coefficients) RAPID/IES data for 2001-2016}
	\label{ratio_small}
\end{sidewaysfigure}

%\newpage
\begin{sidewaysfigure}
	\centering
	\includegraphics[width=\textwidth]{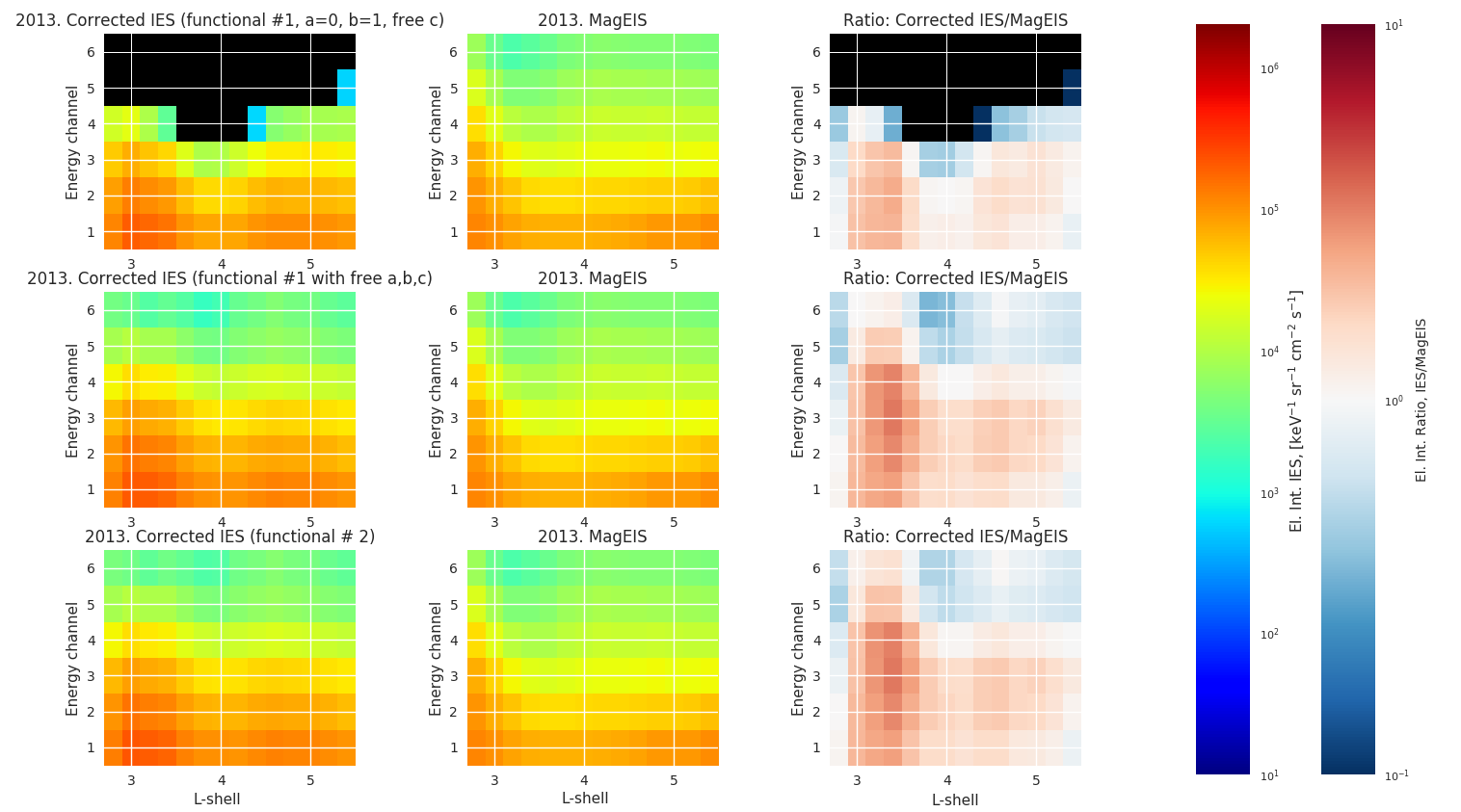}
	\caption{Different functional constructions for RAPID data correction for 2013}
	\label{2013_differential_evolution}
\end{sidewaysfigure}

%\newpage
\begin{sidewaysfigure}
	\centering
	\includegraphics[width=\textwidth]{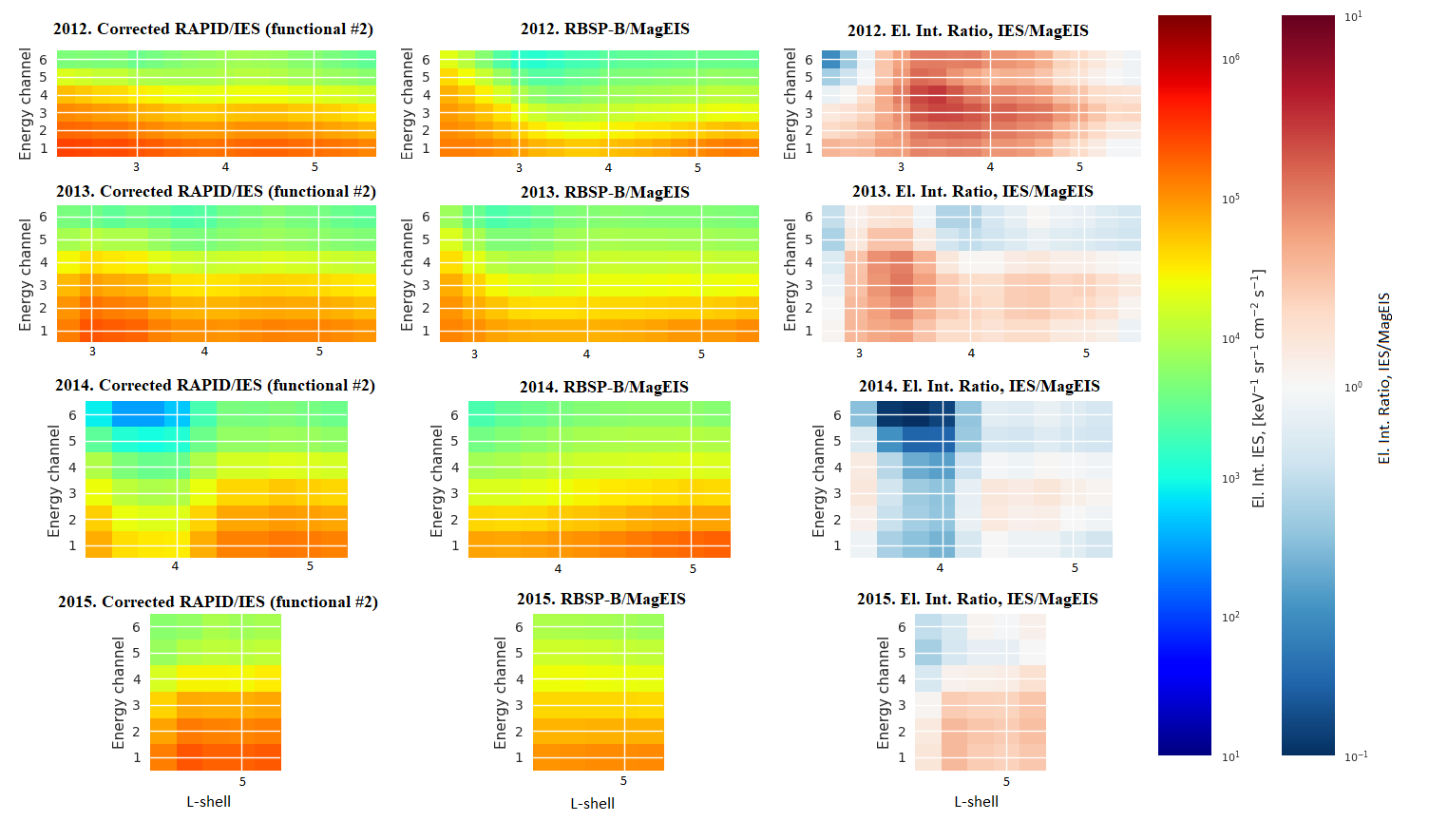}
	\caption{IES electron intensities, corrected using MagEIS proton measurements, corrected RBSP-B electron intensities, and their ratios for 2012-2015, plotted as energy versus L-shells}
	\label{functionals_2012_2015}
\end{sidewaysfigure}

%\newpage
\begin{figure}
	\centering
	\includegraphics[width=\textwidth]{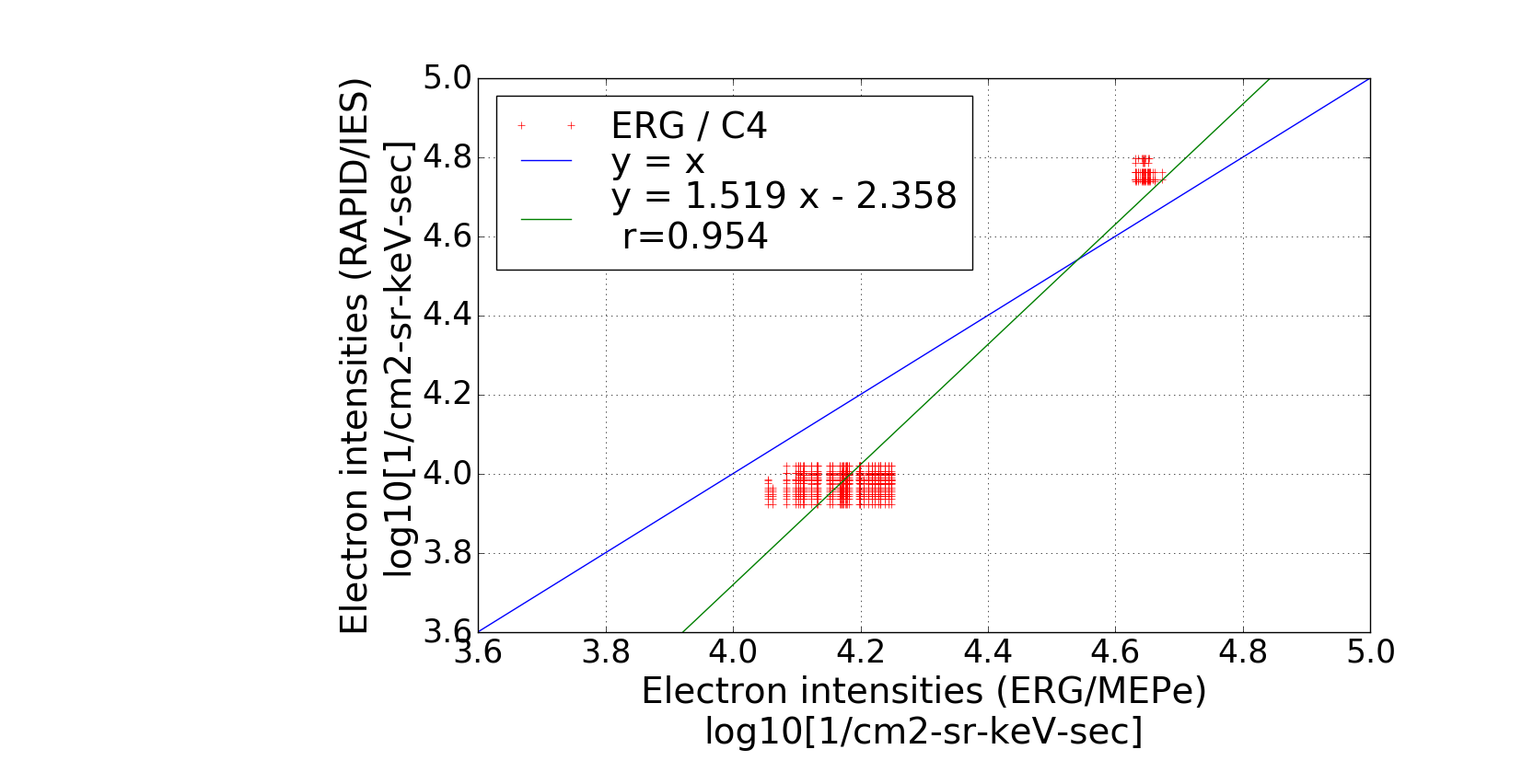}
	\caption{Linear regression of electron intensities measured by ERG and Cluster (spacecraft 4) missions}
	\label{ERG}
\end{figure}

%\newpage
\begin{sidewaysfigure}
	\centering
	\includegraphics[width=\textwidth]{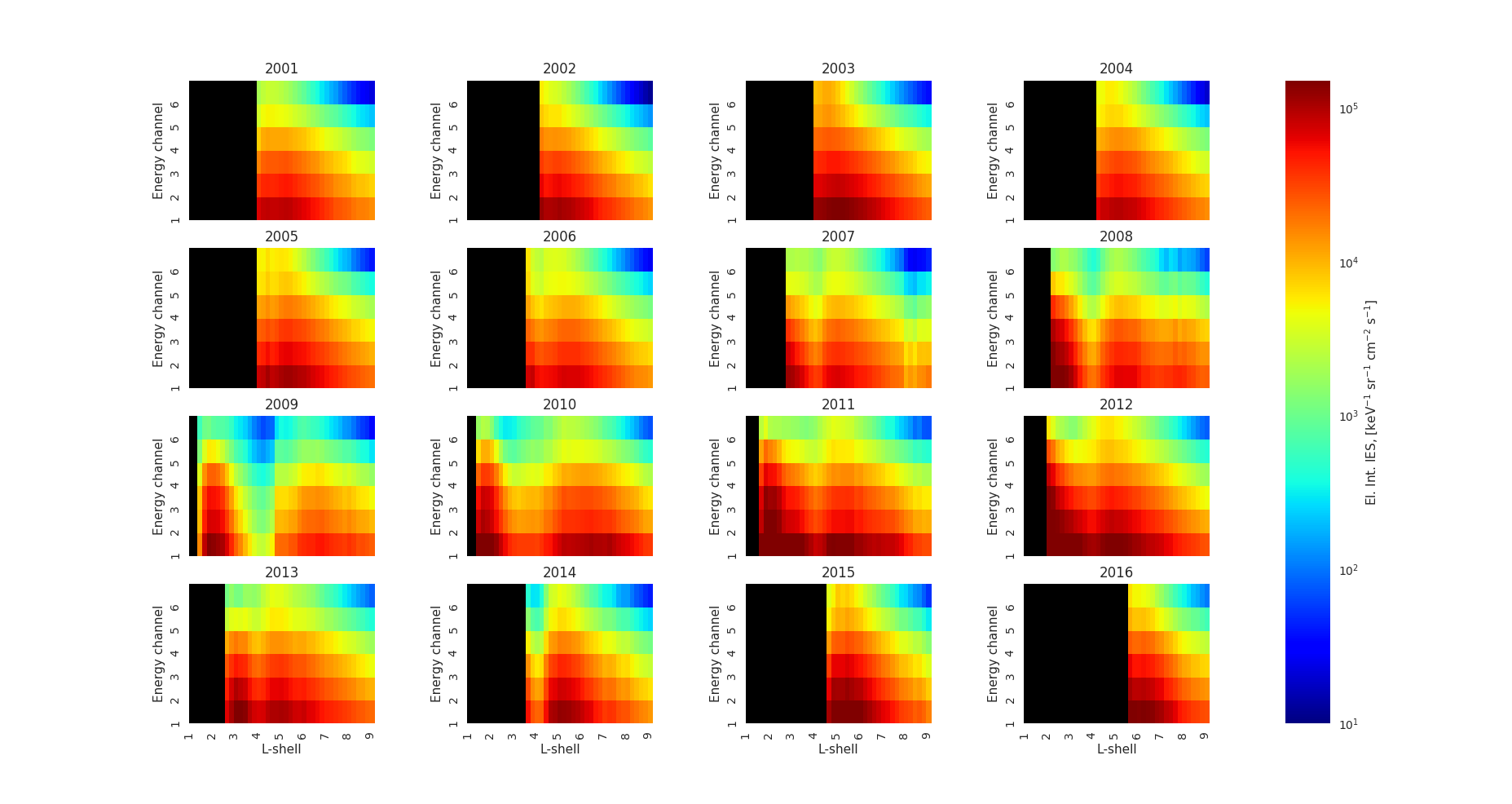}
	\caption{Spin-averaged yearly mean IES electron intensities for 2001-2016 corrected using contamination coefficients from Table 3}
	\label{solar_cycle_electrons}
\end{sidewaysfigure}

%\newpage
\begin{figure}
	\centering
	\includegraphics[width=\textwidth]{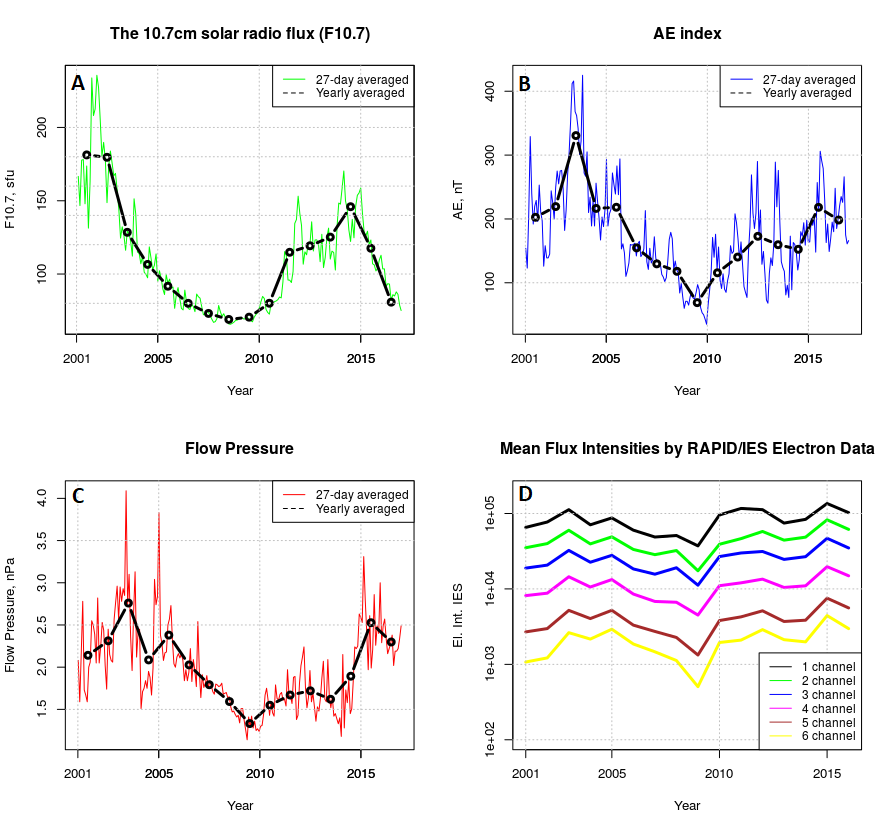}
	\caption{F10.7 solar activity index (A), AE index (B), Dynamic Flow Pressure (C), and Yearly mean flux intensities by IES electron data (D) for 2001-2016}
	\label{indices}
\end{figure} 

%\newpage
\begin{figure}
	\centering
	\includegraphics[width=\textwidth]{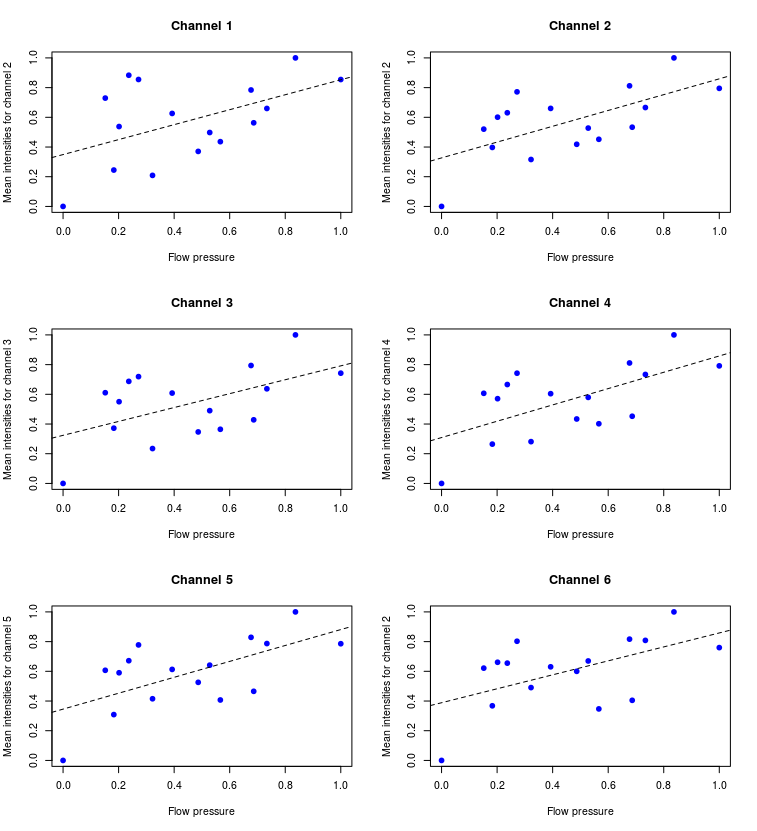}
	\caption{Linear regression analysis as solar wind dynamic pressure versus electron flux intensities for Energy channels 1-6}
	\label{linear_regression}
\end{figure}

%TABLES:
\newpage
\begin{table}[ht]
	\textbf{\caption{Lower Thresholds of 6 Electrons Energy Channels for RAPID}}
	\centering
	\begin{tabular}{c c}
		\hline\hline
		Channel & Electrons (keV) \\ [0.5ex]
		\hline
		1 & 39.2 \\
		2 & 50.5 \\
		3 & 68.1 \\
		4 & 94.5 \\
		5 & 127.5 \\
		6 & 244.1 \\ 
		Upper & 406.5 \\[1ex]
		\hline
		
	\end{tabular}
	\label{lower_thresholds_ies}
\end{table}

\newpage
\begin{table}[ht]
	\textbf{\caption{Lower Thresholds of MagEIS Electrons Energy Channels Used in This Study. Note that the MagEIS channels were time-varying before 08.03.2013; the differences in thresholds are insignificant for statistical purposes}}
	\centering
	\begin{tabular}{c c}
		\hline\hline
		Channel & RBSPB Electrons (keV) \\ [0.5ex]
		\hline
		0 &  32 \\
		1 &  54 \\
		2 &  80 \\
		3 &  108 \\
		4 &  144 \\
		5 &  183 \\
		6 &  226 \\ 
		7 & 232 \\ 
		8 &  350 \\ [1ex]
		\hline
	\end{tabular}
	\label{lower_thresholds_rbsp}
\end{table}	 

\newpage
\begin{sidewaystable}[ht]
	\textbf{\caption{Total contamination (\%) of the IES Detector in the Earth's Radiation Belts [after Kronberg et al., 2016] }}
	\centering
	\begin{tabular}{c c c c c c c c c c}
		\hline\hline
		Channel & $L^{\star}=1$ & $L^{\star}=2$ & $L^{\star}=3$ & $L^{\star}=4$ & $L^{\star}=5$ & $L^{\star}=6$ & $L^{\star}=7$ & $L^{\star}=8$ & $L^{\star}=9$ \\ [0.5ex]
		\hline
		1 & 2.63 & 2.87 & 19.59 & 31.41 & 9.27 & 1.33 & 0.44 & 0.26 & 0.17 \\
		2 & 7.51 & 7.95 & 32.96 & 41.08 & 15.09 & 3.24 & 1.3 & 0.8 & 0.52 \\
		3 & 12.29 & 12.04 & 43.18 & 48.7 & 20.2 & 6.15 & 2.68 & 1.61 & 1.06 \\
		4 & 31.06 & 23.68 & 57.64 & 60.66 & 42.16 & 21.44 & 11.34 & 7.07 & 4.9 \\
		5 & 41.47 & 37.1 & 68.73 & 64.38 & 49.94 & 31.45 & 21.81 & 16.53 & 13.31 \\
		6 & 43.51 & 46.53 & 77.93 & 64.79 & 46.41 & 34.55 & 27.94 & 24.57 & 21.3 \\
		[1ex]
		\hline
	\end{tabular}
	\label{contamination_coefficients}
\end{sidewaystable}

\newpage
\begin{table}[ht]
	\textbf{\caption{Mean and median ratio values between RAPID/IES and RBSP-B electron measurements}}
	\centering
	\begin{tabular}{c c c}
		\hline\hline
		
		Year& Median & Mean\\ [0.5ex]
		\hline
		2012& 1.0& 1.33\\
		2013& 1.0& 0.98\\
		2014& 1.0& 0.90\\
		2015& 1.0& 1.02\\
		\hline
	\end{tabular}
	\label{ratio_ies_rbsp}
\end{table}	 

\newpage
\begin{table}[ht]
	\textbf{\caption{Comparison of the used optimization techniques for 2013}}
	\centering
	\begin{tabular}{c c c}
		\hline\hline
		
		Func. 1, free c& Func. 1, free a,b,c & Func. 2, free a,b,c\\ [0.5ex]
		\hline
		a=0& a=0.11& a=0.10\\
		b=1& b=1& b=1\\
		c=1& c=0.14& c=0.08\\
		f=0.02& f=0.13& f=1.475\\
		p=0.668& p=1.37& p=1.42\\
		m=0.98& m=1.24& m=1.30\\
		\hline
	\end{tabular}
	\label{two_methods_2013}
\end{table}

\newpage
\begin{table}[ht]
	\textbf{\caption{Mean and median values of IES/RBSP-B electron intensities ratio and optimization parameters a, b, c for functional (II)}}
	\centering
	\begin{tabular}{c c c c c c}
		\hline\hline
		
		Year& Mean ratio& Median ratio& a& b& c\\ [0.5ex]
		\hline
		2012& 2.35& 2.33& 0.02& 0.99& 0.09\\
		2013& 1.42& 1.3&0.08& 1& 0.09\\
		2014& 0.72& 0.73& 0.13& 1& 0.06\\
		2015& 1.32& 1.22& 0.14& 1& 0.02\\
		\hline
	\end{tabular}
	\label{mean_median_abc}
\end{table}

\newpage
\begin{table}[ht]
	\textbf{\caption{Maximal and minimal values of flow dynamic pressure and mean electron intensities for energy channels 1-6 in 2001-2016}}
	\centering
	\begin{tabular}{c c c c c}
		\hline\hline
		
		Energy Channel& $I_{min}$& $I_{max}$& $P_{min}$& $P_{max}$\\ [0.5ex]
		\hline
		Channel 1& 4.57& 5.13& 1.13& 2.76\\
		Channel 2& 4.24& 4.92& 1.13& 2.76\\
		Channel 3& 4.05& 4.67& 1.13& 2.76\\
		Channel 4& 3.65& 4.29& 1.13& 2.76\\
		Channel 5& 3.12& 3.88& 1.13& 2.76\\
		Channel 6& 2.70& 3.65& 1.13& 2.76\\
		\hline
	\end{tabular}
	\label{min_and_max}
\end{table}

\newpage
\begin{table}[ht]
	\textbf{\caption{Results of statistical analysis}}
	\centering
	\begin{tabular}{c c c c c c}
		\hline\hline
		
		Energy Channel& Intercept& Int. Error& Slope& Slope Error& R\\ [0.5ex]
		\hline
		Channel 1& 0.35&0.1& 0.51&0.20& 0.51\\
		Channel 2& 0.33&0.1& 0.53&0.17& 0.64\\
		Channel 3& 0.33&0.1& 0.47&0.19& 0.54\\
		Channel 4& 0.31&0.1& 0.55&0.18& 0.62\\
		Channel 5& 0.35&0.1& 0.54&0.18& 0.63\\
		Channel 6& 0.38&0.1& 0.47&0.19& 0.55\\
		\hline
	\end{tabular}
	\label{statistics_result}
\end{table}

\newpage
\begin{table}[ht]
	\centering
	\textbf{\caption{Linear trend equations for IES electron intensities for energy channels 1-6}}
	\begin{tabular}{c c}
		\hline\hline
		
		Energy Channel& Linear trend equation\\ [0.5ex]
		\hline
		
		Channel 1& $\log_{10}\textbf{y}=\textbf{0.35}(\pm0.1)\textbf{+0.51}(\pm0.20)\textbf{x}$\\
		Channel 2& $\log_{10}\textbf{y}=\textbf{0.33}(\pm0.1)\textbf{+0.53}(\pm0.17)\textbf{x}$\\
		Channel 3& $\log_{10}\textbf{y}=\textbf{0.33}(\pm0.1)\textbf{+0.47}(\pm0.19)\textbf{x}$\\
		Channel 2& $\log_{10}\textbf{y}=\textbf{0.31}(\pm0.1)\textbf{+0.55}(\pm0.18)\textbf{x}$\\
		Channel 3& $\log_{10}\textbf{y}=\textbf{0.35}(\pm0.1)\textbf{+0.54}(\pm0.18)\textbf{x}$\\
		Channel 6& $\log_{10}\textbf{y}=\textbf{0.38}(\pm0.1)\textbf{+0.47}(\pm0.19)\textbf{x}$\\
		
		%\hline
	\end{tabular}
	\label{trend_equations}
\end{table}

%%%
\listofchanges
%%%

\end{document}